\documentclass{IOS-Book-Article}

\usepackage{mathptmx}
\usepackage{url}
\usepackage{graphicx}

\def\hb{\hbox to 10.7 cm{}}

\begin{document}

\pagestyle{headings}
\def\thepage{}

\begin{frontmatter}              % The preamble begins here.

%\pretitle{Pretitle}
\title{Toward Enabling Reproducibility for Data-Intensive Research using the Whole Tale Platform}

\markboth{}{September 2019\hb}
%\subtitle{Subtitle}

\author[A]{\fnms{Kyle} \snm{Chard}}
,
\author[B]{\fnms{Niall} \snm{Gaffney}}
,
\author[A]{\fnms{Mihael} \snm{Hategan}}
,
\author[C]{\fnms{Kacper} \snm{Kowalik}}
,
\author[D]{\fnms{Bertram} \snm{Lud\"{a}scher}}
,
\author[D]{\fnms{Timothy} \snm{McPhillips}}
,
\author[E]{\fnms{Jarek} \snm{Nabrzyski}}
,
\author[D]{\fnms{Victoria} \snm{Stodden}%
\thanks{Corresponding Author. E-mail: vcs@stodden.net. This material is based upon work supported by the National Science Foundation under Grant No. 1541450.}},
\author[E]{\fnms{Ian} \snm{Taylor}}
,
\author[F]{\fnms{Thomas} \snm{Thelen}}
,
\author[D]{\fnms{Matthew J.} \snm{Turk}}
and
\author[C]{\fnms{Craig} \snm{Willis}}

\runningauthor{K. Chard et al.}

\address[A]{University of Chicago}
\address[B]{Texas Advanced Computing Center}
\address[C]{National Center for Supercomputing Applications}
\address[D]{University of Illinois at Urbana-Champaign}
\address[E]{University of Notre Dame}
\address[F]{University of California at Santa Barbara}


\begin{abstract}

%\noindent {\bf Abstract}\\
%\\

Whole Tale \url{http://wholetale.org} is a web-based, open-source platform for reproducible research supporting the creation, sharing, execution, and verification of ``Tales'' for the scientific research community. Tales are executable research objects that capture the code, data, and environment along with narrative and workflow information needed to re-create computational results from scientific studies. Creating reproducible research objects that enable reproducibility, transparency, and re-execution for computational experiments requiring significant compute resources or utilizing massive data is an especially challenging open problem. We describe opportunities, challenges, and solutions to facilitating reproducibility for data-and compute-intensive research, that we call ``Tales at Scale,'' using the Whole Tale computing platform. We highlight challenges and solutions in frontend responsiveness needs, gaps in current middleware design and implementation, network restrictions, containerization, and data access. Finally, we discuss challenges in packaging computational experiment implementations for portable data-intensive Tales and outline future work.
\end{abstract}

\end{frontmatter}

%\end{frontmatter}
\markboth{September 2019\hb}{September 2019\hb}

\section{Introduction}

In this work we explore barriers and opportunities for extending the Whole Tale infrastructure \cite{Wholetale2019} to facilitate reproducible data-intensive research at scale. Creating reproducible research objects that capture artifacts such as data, software, and sufficient details from a computational experiment to enable reproducibility, transparency, and re-execution is a challenge the research community is addressing in a variety of ways \cite{Donoho2009, stodden-default-reproducible-2013, collberg-repeatability-2016, popper2016, sciunit, reprozip, clusterjob, cranmer-recast-2011, sandve-ten-simple-rules-2013}. However reproducing results that require significant compute resources, rely on specialized hardware, or utilize massive data, is an especially challenging open problem. We start by describing the Renaissance Simulations \cite{Shea2015} to provide a concrete motivating scenario for this work. We then present the current implementation of the Whole Tale reproducible research platform, and define the notion of a ``Tale'' as a reproducible research object published by the Whole Tale system \cite{WholeTaleTale2019, Chard2019a}. We enumerate possible execution models for data-intensive computational experiments that rely on significant compute resources, specialized hardware, and/or massive data. We discuss challenges and solutions inherent in extending the Whole Tale platform this way, including frontend responsiveness needs, gaps in middleware design and implementation, network restrictions, containerization, and data access. We close with a discussion of the role of computational reproducibility in the search for scientific correctness.

We acknowledge the following contributions to this work, following the CASRAI CRediT (Contributor Roles Taxonomy) convention (see \url{https://casrai.org/credit/}). Conceptualization: Chard, Gaffney, Hategan, Kowalik, Willis; Funding acquisition: Lud\"{a}scher, Stodden, Turk, Chard, Nabrzyski, Gaffney; Project management: Kowalik, Willis; Software: Hategan, McPhillips, Kowalik, Taylor, Thelen, Willis; Supervision: Chard, Gaffney, Kowalik, Lud\"{a}scher, Nabrzyski, Stodden, Turk, Willis; Visualization: Hategan; Writing (original draft): Hategan, Stodden; Writing (review \& editing): Chard, Lud\"{a}scher, Hategan, Stodden, Kowalik, Willis.

\section{Motivating Scenario: Renaissance Simulations Laboratory}

To motivate the discussion, we describe a real-world scenario based on the Renaissance Simulations. The Renaissance Simulations (RS) are some of the largest and most detailed simulations of the formation and evolution of the first galaxies. Created after three years using more than 100 million core hours, the resulting simulations enable the exploration of a variety of research questions concerning structure formation and chemical evolution in the early universe. However, the complexity, depth, and size of these simulations require researchers to have access to specialized resources for analysis. The Renaissance Simulations Laboratory (RSL) is a virtual laboratory devoted to providing access to over 70 TB of raw data and derived data products including halo catalogs, merger trees, and mock observations. RSL exposes data available on systems at the San Diego Supercomputing Center (SDSC) via Jupyter web-based interactive environments and, in a later phase of the project, enable launching jobs on SDSC Comet and other XSEDE resources \cite{Jupyter}. Analyses conducted via RSL are intended to be shared and published to foster a collaborative research community in keeping with the long history of openness and transparency of computational research artifacts in computational astronomy.

This scenario illustrates three key challenges related to our vision of ``Tales at Scale,'' which we define as executable research objects that capture the code, data, and environment along with narrative and workflow information needed to re-create computational results from data- and compute-intensive scientific studies. Creating such reproducible research objects to enable reproducibility, transparency, and re-execution for computational experiments requiring significant compute resources or utilizing massive data is an especially challenging open problem. We describe opportunities, challenges, and solutions to facilitating reproducibility for data-and compute-intensive research using the Whole Tale computing platform. First, the RS data are very large, impractical to transfer, and requires large-scale resources to analyze. Second, the research community leverages interactive Jupyter environments for both exploratory and primary analytical work with some analysis requiring batch compute resources. Third, the community is interested in sharing resulting research artifacts (e.g. code, derived data) for both re-execution and re-use. There are many research communities that would benefit from general-purpose infrastructure, like RSL, that enables researchers to 1) perform exploratory work via large-scale interactive environments, 2) publish reproducible research artifacts based on experiments that require HPC workloads, and 3) do both in an environment that does not require transferring data to new systems. RSL and the Whole Tale project share common platform components and Whole Tale is using the RSL as a driver for the design and implementation of solutions to common problems of computational reproducibility at scale.

\section{The Whole Tale Project: Goals, Infrastructure, and Tales}

In related work we have presented the Whole Tale project: a web-based and open source cyberinfrastructure platform that enables the generation and publishing of reproducible research artifacts, which we call ``Tales,'' objects encapsulating code, data, and computational environment information \cite{WholeTaleTale2019, Chard2019a}. The goal of the Whole Tale project is to enable reproducibility for computation- and data-enabled research. The approach is to transform the discovery process by uniting data products, computational pipelines, and research articles into an integrated whole. Whole Tale supports research experiments in situ, through popular analysis environments such as Jupyter and RStudio interfaces, and captures information including the code, data, provenance, and execution environment used to produce research findings. This information is then packaged by Whole Tale into an archival format called a Tale which defines a standard for executable and reproducible research objects \cite{Chard2019b}. The Tale stores explicit references to data and code used in computational experiments, both for reproducibility purposes and to permit the citation of the specific versions used in any subsequent research. A Tale can be submitted or published to an external research repository and assigned a persistent identifier by the repository. The Whole Tale platform allows users to interactively create and edit Tales and to re-run a Tale to reproduce and verify results as obtained by the original Tale creator. 

The Whole Tale platform itself consists of a set of services, collectively termed the Whole Tale Services, which serve the main web interface and implement the backend services needed to support the functionality of the web interface. Whole Tale is developed as an open source project and can be deployed by third parties. The Whole Tale Services at \url{https:\\wholetale.org}are currently deployed on persistent resources that take the form of a number of virtual machines supporting the underlying databases, service containers, and other components. Currently, research in the Whole Tale environment is supported by a Docker swarm cluster on Jetstream virtual machines \cite{Jetstream, xsede}, referred to at the ``Whole Tale cluster.'' The Whole Tale Services are responsible for launching and managing Tales. Tale ``Frontends,'' such as a Jupyter or RStudio notebook, run on Tale Frontend Resources, which may be co-located with Whole Tale resources.  Tales that require heavy computational or data resources, the focus of this article, are considered to contain HPC Workloads, which typically consists of CPU-intensive serial or parallel (e.g., MPI) code, or code meant to be run on GPU hardware. The lifecycle management of Tale frontends and eventually High Performance Computing (HPC) workloads is handled by specialized middleware, discussed from the perspective of ``Tales at Scale'' will be presented in Section \ref{middleware}.
The Whole Tale platform today focuses on interactive environments as Tale Frontends for both exploratory work and re-execution of published artifacts. Through this, system users are able to work in environments that they are comfortable with while gaining the benefits of the Whole Tale platform to package the results of their work. These interactive environments also provide the opportunity to instrument parts of the research process, such as software dependency identification and computational provenance.  As discussed below, the focus on interactivity can be at odds with support for HPC resources where the primary interactive environment is typically a secure shell such as SSH.

\section{Possible Execution Models Allowing Access to External Resources}

A visual representation of the conceptual architecture described in the previous section is shown in Figure 1. In the current Whole Tale implementation, the three types of resources shown in Figure 1 are only logically distinct and currently consist of Docker containers. The middleware connecting the core services with the Tale frontends is a wrapper around Docker API calls. Tales, in turn, can launch applications that are pre-built into the Tale images using standard system calls, thereby running them inside the same container as the Tale. However, CPU-intensive applications are, in principle, still limited to being run inside the Tale containers. This limitation can be worked around by connecting to external resources and launch custom HPC workloads. Such a solution, however, hinders the ability of a Tale to capture the entirety of the environment in which it was created, negatively impacting reproducibility. This issue will be discussed in more detail in Section \ref{challenges}.

\begin{figure*}
\centering
    \includegraphics[width=\textwidth]{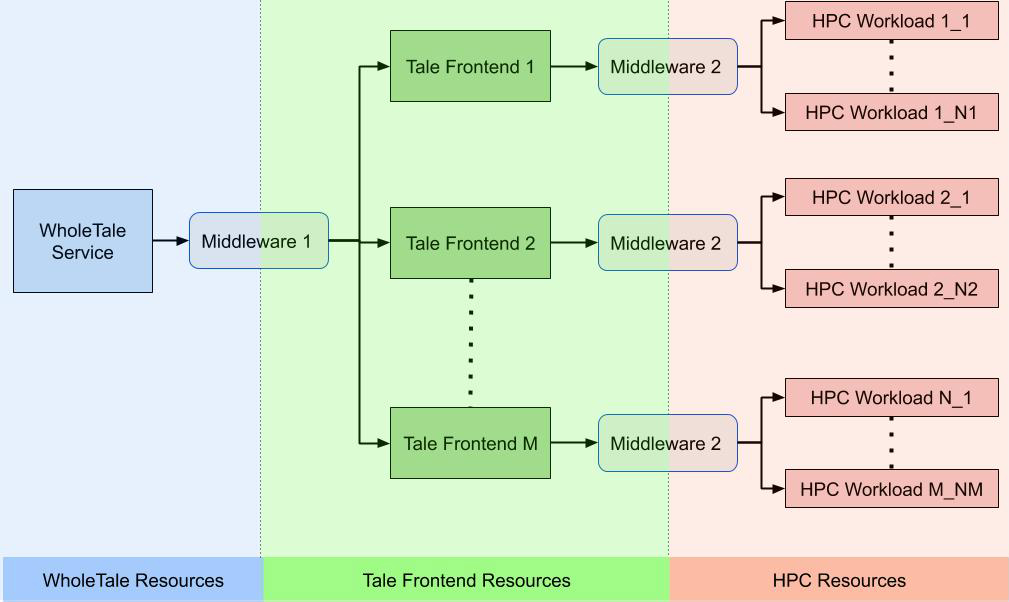}
    \caption{Conceptual high level Whole Tale architecture for executing Tales on HPC resources.}
    \label{fig:overview}
\end{figure*}

The architecture presented in the previous section is general in that it does not specify the precise meaning of the resources involved. The Whole Tale platform only dictates that the core services be deployed on some persistent resources. We can distinguish a number of models that refine this general architecture by assigning concrete resources to the logical ones, enumerated below. We briefly list some of their benefits and downsides, with a more detailed analysis following in Section \ref{challenges}. We discuss six possible approaches in turn.

The first model deploys the Tale frontend and HPC workloads on the Whole Tale cluster (Figure 2). The frontend runs on resources that are part of the Whole Tale deployment and users can launch local HPC jobs using standard system calls, jobs which would run inside the same container as the Tale. The resources available for the HPC jobs are limited to what is provided by default through the Whole Tale deployment cluster.

\begin{figure*}
\centering
    \includegraphics[width=\textwidth]{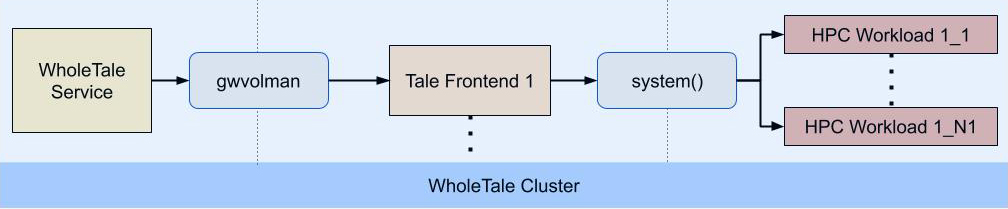}
    \caption{Tale Frontend on Whole Tale Deployment Cluster. The middleware used by the core Whole Tale services to manage Tales is named ``gwvolman.''}
    \label{fig:model1}
\end{figure*}

The second possible model deploys the Tale frontend on an HPC compute node (Figure 3). This option involves running the Tale frontend e.g. a Jupyter notebook or R-studio IDE, on compute nodes in an HPC cluster. The notebooks would then be able to launch independent HPC jobs using standard system calls. As in the previous model, HPC jobs would be local to Tale frontends, but can now benefit from the HPC hardware on which the frontends run.

\begin{figure*}
\centering
    \includegraphics[width=\textwidth]{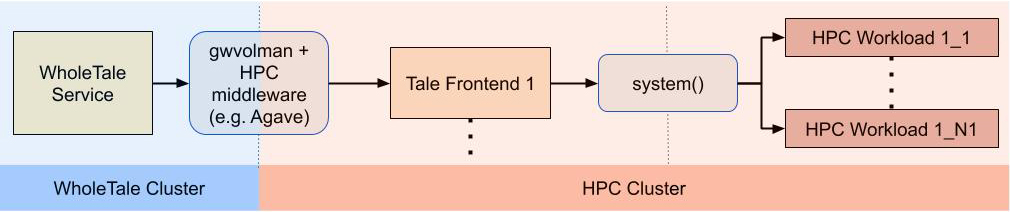}
    \caption{Tale Frontend on single HPC Compute Node.}
    \label{fig:model2}
\end{figure*}

Our third model deploys the Tale frontend on an HPC compute node with local LRM (cluster queuing system) access (Figure 4). This is a similar scenario to that shown in Figure 3, but would allow submission of HPC jobs to the queuing system of the cluster. This would enable scaling of the HPC jobs beyond what is provided by the compute resources available to the Tale frontend.

\begin{figure*}
\centering
    \includegraphics[width=\textwidth]{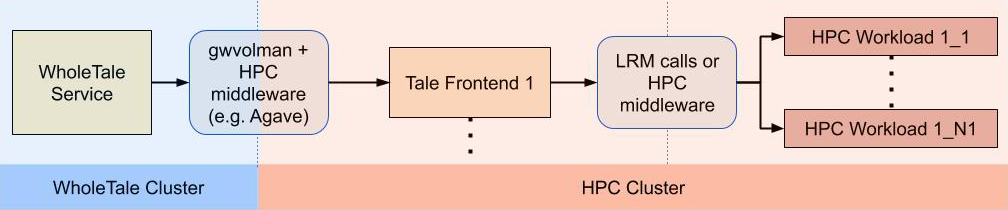}
    \caption{Tale Frontend on HPC Compute Node with Local LRM (cluster queuing system) Access.}
    \label{fig:model3}
\end{figure*}

We can extend the third model to deploy the Tale frontend on HPC compute nodes with MPI (Figure 5). This involves launching the Tale frontend as an MPI job. The cluster LRM (queuing system) would allocate the number of nodes requested at the submission of the Tale frontend job and set the appropriate MPI environment. The Tale frontend would run on the lead node allocated to the MPI job by the LRM and would be able to, using ``mpirun'' or other cluster specific tools, launch MPI subjobs on the nodes allocated to the MPI job.

\begin{figure*}
\centering
    \includegraphics[width=\textwidth]{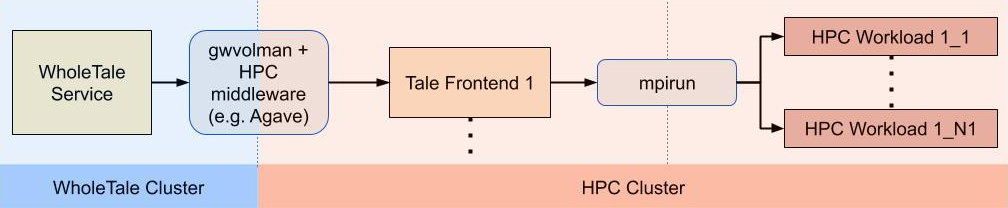}
    \caption{Tale Frontend on HPC Compute Nodes with MPI.}
    \label{fig:model4}
\end{figure*}

Our penutimate model deploys the Tale frontend on the Whole Tale cluster with remote LRM access (Figure 6). In this scenario, Tale frontends continue to run alongside Whole Tale core services, but HPC jobs can be submitted to remote clusters via the middleware. 

\begin{figure*}
\centering
    \includegraphics[width=\textwidth]{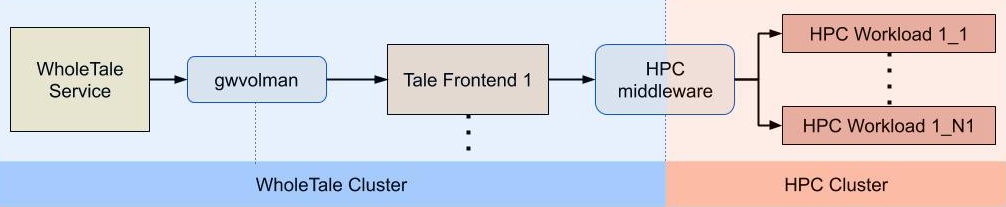}
    \caption{Tale Frontend on Whole Tale Cluster with Remote LRM Access.}
    \label{fig:model5}
\end{figure*}

The final model is a decoupled Tale frontend with LRM Remote Access. In this model, Tale frontends are allowed to run on various resources, including the Whole Tale cluster, a cloud provider, or an HPC cluster. HPC jobs, in turn, can run on any resources supported by the middleware. This model would be useful in allowing users to bypass the limitations present in the default resources provided by the Whole Tale infrastructure. A user with cloud access could request that a Tale be run on cloud resources under the user's account. Furthermore, users with access to HPC clusters could schedule HPC jobs on those resources.

\section{Infrastructure Challenges for Implementing ``Tales at Scale'' on Whole Tale} \label{challenges}

In this section we consider challenges to consider when implementing the ``Tales at Scale,'' model including frontend responsiveness, HPC network restrictions, and containerization.

\subsection{Maintaining Responsiveness of Tale Frontends }

From a usability perspective, Tale frontends need to be launched relatively quickly and predictably. When running Tale frontends on the Whole Tale cluster, Tale frontends become available in a matter of seconds. The process of building the Tale image and launching the container is not instant, but roughly equal to the amount of time it takes docker to load the Tale image from a local registry. However, if the Tale frontend is launched on a typical HPC resource, the process involves the additional step of submitting a job to the LRM and waiting for the LRM to schedule the job on compute nodes. The nature of HPC workloads (often long-running processes) means that LRMs are rarely optimized for quickly scheduling jobs. In addition, the scheduling time is also affected by the number and size of jobs belonging to other users that are already queued by the LRMs. HPC resources may also cycle through scheduled maintenance making them intermittently unavailable. These factors can make the Tale launching time on HPC resources unpredictable, and measured very rarely in seconds.

Strategies to mitigate Frontend launch responsiveness include advanced reservations and pilot jobs. Advanced reservations must be negotiated with HPC resource owners ahead of time. While it may be possible to negotiate long term advanced reservations, this cannot usually be done in an automatic fashion and is repeated whenever a new HPC resource is used. By contrast, an approach based on pilot jobs would involve maintaining a dynamic pool of placeholder HPC workloads that are already scheduled on HPC resources and can pick up actual HPC workloads instantly. From the responsiveness standpoint, solutions which place Tale frontends on the Whole Tale cluster are, therefore, preferred.

\subsection{Scalability and Longevity of Middleware} \label{middleware}

Middleware that supports the ``Tales at Scale'' idea described in this article will allow the Whole Tale core services to launch Tales, and users to launch HPC jobs from Tales. Some examples of middleware include the Globus Toolkit \cite{Foster2006}, ``Simple API for Grid Applications'' (SAGA) \cite{SAGA}, DRMAA (see Global Grid Forum \url{http://www.gridforum.org/}), and Agave (see The Agave Platform \url{http://developer.agaveapi.co/}). Most current middleware provides some level of abstraction over individual HPC resources. That is, to the programmer using the middleware, different resources are exposed through a unified API which abstracts away details such as the type of LRM that a specific HPC resource employs. The nature of the application using the middleware as well as the HPC resources will impose certain constraints on middleware implementation. There are two important issues to consider in our Tales at Scale case: middleware scalability and longevity. 

A problem arises from the fact that scalability may not have been explicitly considered as an integral part of remote HPC submission library design. Such libraries may be designed and tested with the idea of a user submitting and monitoring a relatively low number of jobs at a time. When used by a high throughput system however, such as a portal or a workflow system, problems can arise. For example middleware that wraps LRM command line tools to monitor jobs, such as ``qstat,'' may only periodically do so, even when an invocation occurs for every job handled by the middleware, if the number of jobs is small. When the number of jobs is large however, the repeated invocation of ``qstat'' can overwhelm the LRM.

An additional issue preventing middleware from scaling is the management of secure connections. Establishing secure network connections requires a number of cryptographic steps which tend to be CPU-bound. If each job operation (submission, status query) requires the establishment of a secure connection and many such operators are performed on a single CPU, such as a single middleware client used by a portal or a service deployed on an HPC resource, the rate at which the operations can be performed can be limited. Addressing this issue requires aggregating operations under connections that are kept alive for some time. This is only useful however if the same middleware client/service instance are used. Associating a different client with each Tale instance can potentially reduce the benefits that would come from the sharing of secure connections, but may be necessary for different authentication credentials.

Finally, the lifetime of a remote job as seen from the client side is fundamentally asynchronous. The process involves doing some work until the job information is transmitted to the remote side, followed by waiting until the job completes or fails. It is convenient in many cases to treat such remote jobs as a synchronous process since it leads to simpler code. However, this approach can be wasteful since it involves allocating a thread for each job, a thread which spends most of its lifetime doing nothing while tying up resources. It may be important to note that both the middleware API and the implementation must be asynchronous in order to derive a scalability benefit. For example, the Python implementation of the SAGA provides an asynchronous API layer implemented as a wrapper around a thread based, synchronous core, which is unsatisfactory.

A second important feature of middleware is longevity. By longevity we mean the ability of middleware to continue to perform its intended purpose. From a reproducibility perspective, we must be particularly sensitive to this. Middleware typically used to access HPC resources can lack long term support and funding. Additionally, the incentives required to support stable software produced in a research environment do not always align with the needs of long term executability and reproducibility. A corollary is that ``*-as-a-service'' solutions tend to be insufficient since not only does the client side, including for example Whole Tale, lack long term support, but so does the service side. Even assuming open source and that the possibility of maintaining the client side component of the middleware exists, the service side may still pose challenges, in particular when deployment is on resources that are not under the control of either Whole Tale or users. For example, TeraGrid, the precursor to XSEDE, allowed HPC job submission to resources using specialized services which are not used by XSEDE. As a consequence, software written to work on TeraGrid using the corresponding client libraries would be unable to function today on XSEDE due to a lack of corresponding services. However, preservation of Tales can enable transparency for HPC workflows, even when re-executability is not possible.

Longevity is a difficult problem but solutions are possible. The Whole Tale project could develop a ``middleware insulation service'' which would allow Tales to program against an API that is fixed in time. The Whole Tale team would then maintain working bindings from the insulation service to current HPC middleware. This requires that the Whole Tale project itself is sufficiently supported in the future, which is not a certainty. Alternatively, HPC middleware could be based on protocols that are well known and likely to endure the test of time. Specifically, SSH is nearly universally available on HPC clusters and if the list of LRM implementations is relatively small with stable interfaces, middleware that uses SSH to connect to HPC resources and invokes the relevant LRM commands would be usable even if the core Whole Tale services were unavailable. With the ability to save Tales in a format that allows them to be downloaded from repositories independently of the existence of a working Whole Tale deployment, a feature which we presently support, there is a potential path toward the long term utility of ``Tales at Scale.'' To increase the potential long term utility of ``Tales at Scale'' HPC runs could be made increasingly transparent through provenance capture via the Whole Tale platform, even when re-executability may not be possible. 

\subsection{Managing HPC Network Restrictions }

Many HPC clusters restrict incoming network connections to compute nodes from outside the cluster. Tale frontends however require incoming network connections in order to expose their user interface. Consequently, a general solution involving Tale frontends on compute nodes requires some form of proxying of connections from the Whole Tale cluster to HPC cluster compute nodes. Restrictions on incoming network connections may likely be a result of local security policies and therefore proxying, even if authenticated, may be seen as an unwelcome circumvention of such policies potentially requiring engagement with HPC resource managers. Many computing centers are now deploying Jupyter environments which opens the possibility of leveraging these resources directly.

\subsection{Containerization and HPC Workloads}

For the purpose of containerization, we can divide HPC workloads into unoptimized applications, optimized applications, and mixed applications. Unoptimized applications refer to various tools that use general CPU instructions and can be compiled (or are interpreted) and run on most types of hardware. Optimized applications are designed to benefit from specialized CPU instructions, such as Streaming SIMD Extensions (SSE) instructions or GPU hardware. Mixed applications consist of unoptimized driving code with optimized cores provided by specialized libraries.

From the standpoint of computational reproducibility, optimization poses a fundamental problem, since it benefits from specialization for particular hardware, whereas the ability to re-run code at different points in time and on different hardware requires abstraction. For mixed and unoptimized applications, containerization can capture the environment and dependencies of the unoptimized parts. On the other hand, when dealing with optimized applications or libraries, recording such code in optimized form leads to a dependence on specific hardware which can affect the ability for the code to be re-run if the specific hardware becomes unavailable. Even without that issue, typical optimized HPC applications are compiled for specific hardware and statically linked against libraries specific to that hardware to improve performance, rendering containers unnecessary. Unfortunately providing only precompiled executables in a Tale introduces opacity in the computation that underlies the research, implying that the relevant source code should always be included in a Tale to enables transparency as well as reproducibility, regardless of whether otherwise reproducible binary code is included or not. In addition to including the source code in a published Tale, several possible choices regarding the inclusion of optimized binary code exist:

\begin{enumerate}
\item Package generic, pre-compiled, statically linked executables with Tales. These Tales will therefore be tied to a specific architecture (e.g. AMD64) and may perform suboptimally on specific HPC resources that share the architecture but not necessarily the fine details of custom instruction sets.
\item Package multiple versions of pre-compiled, statically linked executables with Tales, one for each target HPC resource. This provides performance optimality, but is tied to a specific set of resources.
\item Package source code and generic libraries required to compile the source code, which may perform suboptimally on specific HPC resources, but is not necessarily tied to a specific architecture.
\item Provide as much ancillary infrastructure to allow on-demand compilation against custom HPC libraries when a Tale is run. This may include compilation on target HPC resources or cross-compilation including possibly copies of libraries/compilers optimized for target HPC resources. 
\end{enumerate}

Each of these choices involves tradeoffs between reproducibility/portability and Tale size and complexity. For example, GCC on BlueGene/Q machines produces unoptimized code that, three years ago, was significantly slower than code produced by the IBM XL compiler which used specialized PPC vector instructions. However, the XL compiler is proprietary and, even if available for free, the license of proprietary tools may or may not permit redistribution in a form suitable for direct inclusion in Tales.

In addition, an instance of an experiment must decide on a concrete set of dependencies. This can lead to an overspecification, for example including irrelevant constraints on the dependencies, such as specific versions of libraries. This can be problematic if repeatability is affected when such constraints are relaxed. To give a fabricated by not atypical example, imagine an experiment uses a linear algebra library to multiply matrices and specifies that version 1.0.2 of the library should be used. When the experiment is re-run with version 1.0.2 of the library, the same results are obtained. However, when version 1.0.3 is used, the results differ. This example shows how altering dependencies can change scientific results obtained at the application level.

\subsection{Data access and Data Quasi-locality}

The current implementation of Whole Tale uses a cache (internally called the Data Management System, or DMS) which brings data from external collections to Whole Tale resources. It does the relevant transfer when users attempt to open the corresponding files from inside a Tale. However, different strategies are available and initiating transfers when a Tale is created is possible. This would result in a quasi locality of data since, after the initial transfer, the data is local. There are limitations to this type of solution. One is that data repositories may contain significantly more data than would be feasible to transfer and store on Whole Tale resources. On the other hand, for many applications only a small subset of the data in external repositories would be actively used at any given time. Data access and usage patterns remain an open question which could inform the viability of particular solutions.

In the event that Tale frontends and/or HPC workloads run on HPC resources on which copies of data are already available, the current Whole Tale implementation would be inefficient, since each file would be transferred once to Whole Tale resources and once for each Tale frontend instance that accesses the file. Bypassing this mechanism requires that we develop an alternative solution to the current DMS. The precise details depend on how such data are exposed on HPC resources. There are two main options. The first occurs when data are accessible on HPC resources from a POSIX file system. In this case, the solution consists simply of instructing the container system, if one is used, to mount the relevant files/directories inside the Tale container. The second option is that data are available on HPC resources through a non-POSIX interface. In this scenario, Tale initialization (or HPC task initialization) must invoke the relevant tools to transfer the files to a locally accessible location on the HPC resource.

\section{Conclusions and Future Work}

We have attempted to describe the challenges inherent in extending the Whole Tale platform to enable reproducible data-intensive research at scale including frontend responsiveness needs, gaps in current middleware design and implementation, network restrictions, containerization, and data access. We have provided several examples outlining possible steps forward for the Whole Tale system in advancing reproducibility for data- and compute-intensive scientific applications, related to work on the adoption of common workflow systems and the capture of detailed experiment provenance information for reproducibility \cite{Galaxy, Kepler, Pegasus, Hunold2013, Fursin2018, Pouchard2019}.

The ``Tale'' specification \cite{Chard2019b} is designed to capture and communicate information needed to reproduce the scientific results it contains (at least contemporaneously), and includes all code, data, libraries, and environment information (collectively, dependencies) necessary for an experiment to be re-run. However, regenerating identical computational results when an experiment is repeated does not necessarily indicate scientific correctness. Rerunning faulty experiments can produce the same faulty results. The task of relaxing dependencies to a necessary minimum is complicated by the fact that non-trivial software may require overspecified dependencies due to reasons that are not relevant to the experiment. For example, libraries may themselves have more or less strict constraints on their dependencies, which are due to code paths that are not exercised by a particular experiment. It is perhaps important to recognize that the role of projects like Whole Tale is primarily one of accurate recorders that remove the guesswork from the task of understanding the context in which the original experimenters obtained their results. By contrast, the task of ensuring scientific correctness is mainly left to the scientific community at large. This suggests that, faced with many technical challenges in supporting the re-executability of complex HPC workloads, a more practical short term goal might be to focus on enabling the systematic capture of provenance information, recording the conditions under which experiments were performed. This instrumentation process would still benefit from access to the interactive environment in which the experiment is run.

Through its creation of ``Tales,'' the Whole Tale platform embraces reproducibility as a form of packaging for transparency. Today, the Whole Tale system can be used to package data- and compute-intensive research artifacts and provide an interactive environment for exploration. The Whole Tale platform could also be extended to enable researchers to launch interactive environments during the exploratory phase of their research as well which may require providing access to specialized compute resources or data. Whole Tale can achieve this goal today in the context of running on a single node. We have also described the more challenging case of capturing information regarding code, data, and the computational environment via the interactive environment provided by Whole Tale, including provenance information \cite{Pouchard2019}, with the goal of enabling re-execution for data-intensive workloads at scale.

\end{document}